\documentclass[twoside,a4paper,11pt]{proceedings}
\usepackage{graphicx}
\usepackage{hyperref}
\usepackage{ifpdf}
\usepackage{movie15}
\usepackage{natbib}
\begin{document}
\pagenumbering{arabic}
\pagestyle{myheadings}
\thispagestyle{empty}
\vspace*{-1cm}
{\flushleft\includegraphics[width=3cm,viewport=0 -30 200 -20]{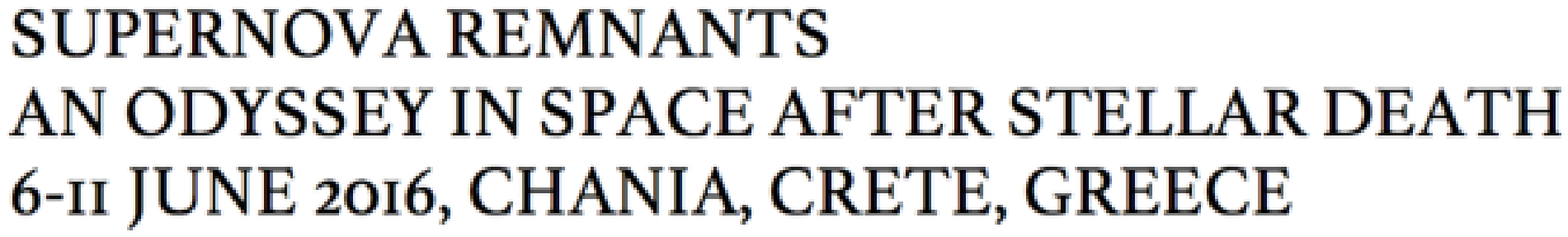}}
\vspace*{0.2cm}
\begin{flushleft}
{\bf {\LARGE
Observations of Supernova Remnants with the Sardinia Radio Telescope
}\\
\vspace*{1cm}
E. Egron$^1$,
A. Pellizzoni$^1$,
S. Loru$^1$,
M. N. Iacolina$^1$,
M. Marongiu$^1$,
S. Righini$^2$,
S. Mulas$^3$,
G. Murtas$^3$,
M. Bachetti$^1$,
R. Concu$^1$,
A. Melis$^1$,
A. Trois$^1$,
R. Ricci$^2$
and M. Pilia$^1$
%
}\\
\vspace*{0.5cm}
%
$^{1}$
INAF, Osservatorio Astronomico di Cagliari, Via della Scienza 5, 09047 Selargius, Italy \\
$^{2}$
INAF, Istituto di Radioastronomia, Via Gobetti 101, 40129 Bologna, Italy \\
$^{3}$
Dipartimento di Fisica, Universita' degli studi di Cagliari, SP Monserrato-Sestu, KM 0.7, 09042 Monserrato, Italy\\
%
\end{flushleft}
\markboth{
Obsevations of Supernova Remnants with SRT
}{
Egron et al.
}
\thispagestyle{empty}
\vspace*{0.4cm}
\begin{minipage}[l]{0.09\textwidth}
\ 
\end{minipage}
\begin{minipage}[r]{0.9\textwidth}
\vspace{1cm}
\section*{Abstract}{\small
In the frame of the Astronomical Validation activities for the 64m 
Sardinia Radio Telescope, we performed $5-22$ GHz 
imaging observations of the complex-morphology supernova remnants (SNRs) W44 and IC443. We adopted innovative observing and mapping techniques providing 
unprecedented accuracy for single-dish imaging of SNRs at these 
frequencies, revealing morphological details typically available only at 
lower frequencies through interferometry observations.
High-frequency studies of SNRs in the radio range are useful to better 
characterize the spatially-resolved spectra and the physical parameters 
of different regions of the SNRs interacting with the ISM. Furthermore, 
synchrotron-emitting electrons in the high-frequency radio band are also 
responsible for the observed high-energy phenomenology as -e.g.- Inverse 
Compton and bremsstrahlung emission components observed in gamma-rays, to 
be disentangled from hadron emission contribution (providing constraints 
on the origin of cosmic rays).

\vspace{10mm}
\normalsize}
\end{minipage}

\section{Introduction}


The Sardinia Radio Telescope (SRT, www.srt.inaf.it) is a new 64-m single-dish antenna operated by INAF (Istituto Nazionale di Astrofisica; Italy). The advanced technology, in particular the active surface, will allow us to observe frequencies from 300 MHz up to 115 GHz. 
We proposed innovative observing and mapping techniques during the Astronomical Validation phase, with the development of the Single Dish Imager (SDI; Pellizzoni et al. in prep.). This software is dedicated to the production of calibrated maps of extended sources, such as SNRs and pulsar wind nebulae. 
\newline We present the imaging of the Galactic SNR IC443 at 7.24 GHz obtained with SRT during the Astronomical Validation phase. We compared our results with high-resolution maps of this source obtained with the VLA and Arecibo at 1.4 GHz (Lee et al. 2008).


\section{SRT observations}

We carried out four observations of IC443 at 7.24 GHz (LO=6800 MHz; bandwidth=680 MHz) from May 27 to December 10 2014.  The data were recorded with the Total-Power backend, an analogic to digital converter.
The active surface was set in the shaped configuration to offer a better illumination of the Gregorian focus and to adjust the panels of the primary mirror in function of the elevation. The minor servo system was configured in tracking mode to correct the sub-reflector position according to its pointing model. 

We performed mapping of IC443 through On-the-Fly scans (OTF). This technique implies that the data acquisition is performed with continuity (sampling time of 40 ms), at constant speed (typically a few degrees/min), repeatedly scanning the sky in both right ascension (RA) and declination (DEC) directions. The subscan length was set to $1.5^{\circ}$ in both RA and DEC, accounting for the size of the target of $\sim 45'$ and baseline subtraction requirements. Each subscan duration was scheduled to 22.5 sec, which implies an OTF speed of $4'$/sec. Two consecutive subscans were separated by an offset of $0.01^{\circ}$, which implies 4.5 passages per beam on average, and about 17 samples per beam per scan (assuming a beam size of $2.66'$ at 7.24 GHz). 
The total duration of a single map (RA+DEC) was $\sim 2.5$ hours.

\section{Results}

Data analysis was performed through the SDI, a tool designed to perform continuum and spectro-polarimetric imaging, optimized for OTF scan mapping, and suitable for all SRT receivers/backends. 
SDI provides an automatic pipeline (quicklook analysis) and interactive tools for data inspection, baseline removal, RFI rejection and image calibration (standard analysis). 
We analyzed the individual maps then the final image resulting in merging the data of IC443 obtained at 7.24 GHz.
%
%

We compared our results with high-resolution observations conducted with the VLA and Arecibo at 1.4 GHz and combined together to achieve an extremely good sensitivity and angular resolution of $\sim 40''$ (Lee et al. 2008). 
The details in the complex morphology of IC443 obtained with SRT at 7.24 GHz are comparable with interferometric observations carried out at lower frequencies, as testified by the Fig.1.
SNR IC443 consists in two nearly concentric shells, presenting a clear difference in the radio continuum intensity. The bulk of the emission comes from the northeastern part of the remnant  represented in red in SRT and VLA/Arecibo maps (see Fig.1). This shell is open on the western side on a second shell, which is much more diffuse (Lee et al. 2008; Mufson et al. 1986; Dickel et al. 1989).




\begin{figure}
\center
\includegraphics[scale=0.45]{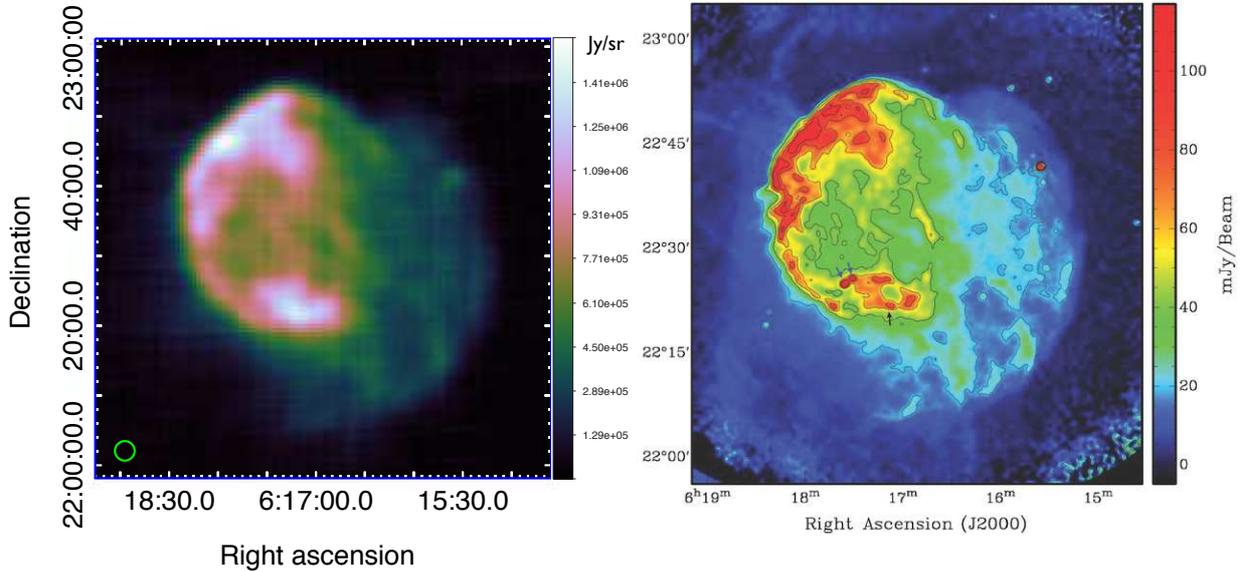} 
\caption{Comparison of the final map of IC443 performed with SRT at 7.24 GHz \textit{(left)} with the combined map obtained with the VLA and Arecibo at 1.4 GHz from Lee et al. 2008 \textit{(right)}.}
\end{figure}



\section{Conclusion}

The first maps of IC443 obtained with SRT at 7.24 GHz are very promising, giving the possibility to study more in details complex sources.
The data analysis was performed using the Single Dish Imager (SDI) software, a new tool that demonstrates the capabilities of SRT in performing single-dish images in C-band. 
SDI will be made available to SRT users in future AO/"Call for Proposals".
The resulting maps provide a detailed structure of the remnants, comparable to interferometric observations carried out with the VLA at lower frequencies (Lee et al. 2008; Castelletti et al. 2011). This testifies the excellent capabilities of SRT in making maps of extended sources using OTF observations. This is of great interest to infer the flux in different resolved regions of sources. 
Further results related to the observations of SNRs IC443 and W44 at 1.4 GHz and 7 GHz are presented in Egron et al. in prep.
%
\newline
\newline
\small  
%
%
\textit{This work is based on commissioning observations with SRT operated by INAF. For observations at SRT, we also credit the Astrophysical Validation Team 
 http://www.srt.inaf.it/astronomers/astrophysical-validation-team/}.

\section*{References}
\bibliographystyle{aj}
\small
\bibliography{proceedings}

Castelletti, G., Dubner, G., Clarke, T. \& Kassim N. E. 2011, A\&A, 534, A21 \\
Dickel, J. R., Williamson, C. E., Mufson, S. L., \& Wood, C. A. 1989, AJ, 98, 1363 \\
Mufson, S. L., McCollough, M. L., Dickel, J. R. et al. 1986, AJ, 92, 1349 \\
Lee, J. J.,  Koo, B. C., Yun, M. S. et al. 2008, AJ, 135, 796 \\
\end{document}